\documentclass[12pt,a4paper]{article}
\usepackage{epsfig}
\usepackage{amsmath,amsfonts,amssymb}
\usepackage{t1enc}
\usepackage{cite}

\begin{document}
\vspace*{-3cm}
\begin{flushright}
hep-ph/0403243 \\
March 2004
\end{flushright}

\begin{center}
\begin{Large}
{\bf CP violation in selectron cascade decays \\[0.2cm]
$\boldsymbol{\tilde e_L \to e \tilde \chi_2^0 \to e \tilde \chi_1^0 \mu^+
\mu^-}$}
\end{Large}

\vspace{0.5cm}
J. A. Aguilar--Saavedra \\[0.2cm] 
{\it Departamento de Física and GTFP, \\
  Instituto Superior Técnico, P-1049-001 Lisboa, Portugal} \\
\end{center}
 
\begin{abstract}
Selectron decays constitute a source of 100\% polarised neutralinos, whose
helicity is fixed by the charge and ``chirality'' of the decaying selectron. In
SUSY scenarios where the second neutralino $\tilde \chi_2^0$ has three-body
decays, the cascade decay
$\tilde e_L \to e \tilde \chi_2^0 \to e \tilde \chi_1^0 \mu^+ \mu^-$ provides
a clean place to study CP violation in the neutralino
sector, through the analysis of CP-violating asymmetries involving the $\tilde
\chi_2^0$ spin $\vec s$ and the momenta of the two muons. We show that a
CP-violating asymmetry in the triple product
$\vec s \cdot (\vec p_{\mu^-} \times \vec p_{\mu^+})$
could be observable at a 800 GeV linear collider provided the gaugino mass $M_1$
has a large phase at the electroweak scale.
\end{abstract}



\section{Introduction}
\label{sec:1}

Supersymmetric theories \cite{susy1,susy2,susy3} are perhaps the best motivated
extensions of the standard model (SM). If
supersymmetry (SUSY) is realised in nature it must be broken, possibly at a
relatively low energy scale.
SUSY-breaking interactions are usually assumed to be flavour-blind, at least
approximately. Otherwise, they would lead to unacceptable rates for 
flavour-changing neutral currents. Analogously, large phases in the
SUSY-breaking terms of the Lagrangian give supersymmetric contributions to
electric dipole moments (EDMs) far above present limits. For a relatively light
SUSY spectrum, the only solutions to overcome this problem are,
either assume that all SUSY-breaking parameters
have very small phases (or are real), or to arrange cancellations among the
different contributions to EDMs to satisfy experimental limits
\cite{pr1,pr2,kane}.

Although the assumption that SUSY-breaking terms are real is conceptually
simpler and more attractive, the possibility of large phases and apparently
``fine-tuned'' internal cancellations must not be
discarded. Indeed, in quantum field theory the Lagrangian parameters are complex
in general, unless there is some argument ({\em e.g.} hermiticity of the
Lagrangian or some symmetry) requiring them to be real. Thus,
setting the phases of SUSY-breaking terms to zero ``by
hand'' may also be regarded as fine tuning, in the absence of a symmetry
principle to explain why these terms must be (approximately) real.
The possibility of large SUSY CP-violating phases
makes compulsory to explore their effects in phenomenology, in order
to find out their presence and determine their magnitude.

In this Letter we are interested in the direct observation of CP violation due
to supersymmetric CP phases, rather than in their effect in CP-conserving
quantities such as cross sections and decay widths (see for instance Refs.
\cite{CPcon1,CPcon2,CPcon3,CPcon4}). We restrict ourselves to the minimal
supersymmetric standard model and focus on the neutralino
sector, studying CP asymmetries in the cascade decay 
$\tilde e_L^\pm \to e^\pm \tilde \chi_2^0 \to e^\pm \tilde \chi_1^0 \mu^+
\mu^-$.
The $\tilde \chi_2^0$ produced in $\tilde e_L^\pm$ decays are 100\% polarised,
having negative helicity in $\tilde e_L^- \to e^- \tilde \chi_2^0$
and positive helicity in $\tilde e_L^+ \to e^+ \tilde \chi_2^0$. Having perfect
$\tilde \chi_2^0$ polarisation is obviously a great advantage for the study of
angular correlations involving the $\tilde \chi_2^0$ spin \cite{note}, and in
particular for the study of CP asymmetries.
Selectrons will be discovered at LHC \cite{paco} if they have masses of a few
hundred GeV, but the large backgrounds present make it impossible a detailed
study of their properties, which must be carried out at a $e^+ e^-$ collider.
Therefore, as source for left selectrons
we consider $\tilde e_L \tilde e_L$ and $\tilde e_R \tilde e_L$
production in the processes
\begin{align}
e^+ e^- & \to \tilde e_L^+ \tilde e_L^- \to e^+ \tilde \chi_1^0 e^- \tilde
  \chi_2^0 \to e^+ \tilde \chi_1^0 e^- \tilde \chi_1^0 \mu^+ \mu^- \,,
  \nonumber \\
e^+ e^- & \to \tilde e_L^+ \tilde e_L^- \to e^+ \tilde \chi_2^0 e^- \tilde
  \chi_1^0 \to e^+ \tilde \chi_1^0 \mu^+ \mu^- e^- \tilde \chi_1^0 \,,
  \nonumber \\
e^+ e^- & \to \tilde e_R^+ \tilde e_L^- \to e^+ \tilde \chi_1^0 e^- \tilde
  \chi_2^0 \to e^+ \tilde \chi_1^0 e^- \tilde \chi_1^0 \mu^+ \mu^- \,,
  \nonumber \\
e^+ e^- & \to \tilde e_R^- \tilde e_L^+ \to e^- \tilde \chi_1^0 e^+ \tilde
  \chi_2^0 \to e^- \tilde \chi_1^0 e^+ \tilde \chi_1^0 \mu^+ \mu^- \,.
\label{ec:proc}
\end{align}
In these processes all final state
momenta can be determined, and the selectron and neutralino rest frames can be
reconstructed \cite{npb}, allowing the study of CP asymmetries involving the
$\tilde \chi_2^0$ spin $\vec s$ and the momenta of $\mu^+$, $\mu^-$ in the
$\tilde \chi_2^0$ rest frame. We do not consider $\tilde e_R^+ \tilde e_R^-$
production with one selectron decaying to $e \tilde \chi_2^0$, because the
branching ratio of this decay is very small in general.
We discuss SUSY scenarios where the second neutralino has three-body decays,
in which case it is possible to have a CP asymmetry in the triple product
$\vec s \cdot (\vec p_{\mu^-} \times \vec p_{\mu^+})$ of order 0.1.
In scenarios with two-body decays $\tilde \chi_2^0 \to \chi_1^0 Z$
the asymmetry is of order 0.02, and when two-body decays to sfermions dominate
it is negligible.
We consider a CM energy $E_\mathrm{CM} = 800$ GeV, as proposed for
a TESLA upgrade. At this CM energy, the cross sections for $\tilde e_L$
production are higher than at 500 GeV due to the smaller destructive
interference between $s$- and $t$-channel contributions and
the larger phase space available.

\section{Decay of $\boldsymbol{\tilde \chi_2^0}$ and CP asymmetries}
\label{sec:2}

\begin{figure}[htb]
\begin{center}
\begin{tabular}{ccc}
\mbox{\epsfig{file=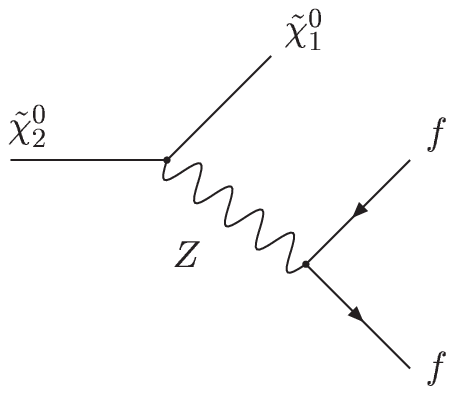,width=3cm,clip=}} & \hspace*{5mm} & 
\mbox{\epsfig{file=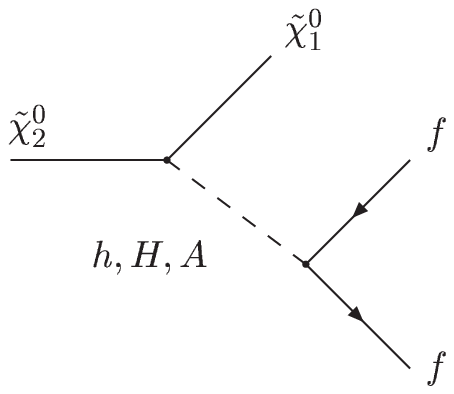,width=3cm,clip=}} \\
(a) & & (b) \\[0.5cm]
\mbox{\epsfig{file=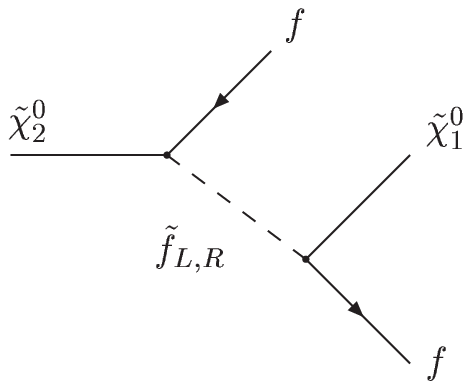,width=3cm,clip=}} & \hspace*{5mm} & 
\mbox{\epsfig{file=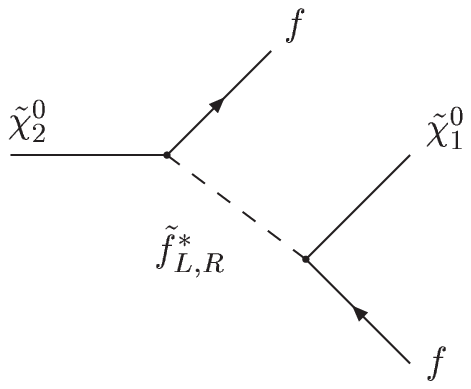,width=3cm,clip=}} \\
(c) & & (d)
\end{tabular}
\caption{Feynman diagrams for the decay $\tilde \chi_2^0 \to \tilde \chi_1^0
f \! \bar f$, mediated by $Z$ bosons (a), neutral Higgs bosons (b),
and scalar fermions (c and d).}
\label{fig:X2decay}
\end{center}
\end{figure}

The decays of the second neutralino $\tilde \chi_2^0 \to \tilde \chi_1^0 f \!
\bar f$ are mediated by the Feynman diagrams in Fig.~\ref{fig:X2decay}.
(The diagrams in Fig.~\ref{fig:X2decay}b with neutral scalars are negligible
except for $f=t,b,\tau$ because they are proportional to the fermion Yukawa
couplings.)
In this work we are interested in the final state with $\bar f \! f = \mu^+
\mu^-$, in which the energies, momenta and charge of both particles can be
measured. The decay channel with $\bar f \! f=e^+ e^-$ shares these
properties, but the multiplicity of electrons in the final state makes
it difficult to identify the ones resulting from the $\tilde \chi_2^0$ decay.
The $\bar b b$ final state is also interesting for its large branching ratio,
but the $b$ tagging efficiency reduces the signal to the same cross
section of the $\mu^+ \mu^-$ channel.

We consider SUSY scenarios where the second neutralino has three-body decays.
This happens when all sleptons are heavier than $\tilde \chi_2^0$ and
$m_{\tilde \chi_2^0}  < M_Z + m_{\tilde \chi_1^0}$. For definiteness,
we choose a scenario similar to SPS1a in Ref.~\cite{sps} but with a
heavier sfermion spectrum and complex phases in $M_1$ and
$\mu$. In this scenario $\tilde \chi_1^0$ and $\tilde \chi_2^0$ are
gaugino-like. The low-energy parameters (at the scale $M_Z$)
most relevant for our analysis are
collected in Table~\ref{tab:1}. For $\phi_1 = \phi_\mu = 0$, these parameters
approximately correspond to $m_{1/2} = 250$ GeV,
$m_{\tilde E} = m_{\tilde L} = m_{H_i} = 200$ GeV, $A_E = -200$ GeV at the
unification
scale, and $\tan \beta = 10$. We use {\tt SPheno} \cite{spheno} to calculate
sparticle masses and mixings, as well as some decay widths.
Neutralino masses slightly depend on $\phi_1$ and
$\phi_\mu$. For $\phi_1 = \phi_\mu = 0$ they are $m_{\tilde \chi_1^0} = 99$ GeV,
$m_{\tilde \chi_2^0} = 178$ GeV, $m_{\tilde \chi_3^0} = 384$ GeV,
$m_{\tilde \chi_4^0} = 400$ GeV, $m_{\tilde \chi_1^-} = 177$ GeV, and for other
values of $\phi_1$ they differ up to $\pm 2$ GeV. The relevant branching ratios
(taking $\phi_1 = \phi_\mu = 0$) are
$\mathrm{Br}(\tilde e_L \to e \, \tilde \chi_1^0) = 0.18$,
$\mathrm{Br}(\tilde e_R \to e \, \tilde \chi_1^0) = 0.998$,
$\mathrm{Br}(\tilde e_L \to e \, \tilde \chi_2^0) = 0.30$,
$\mathrm{Br}(\tilde \chi_2^0 \to \tilde \chi_1^0 \mu^+ \mu^-) = 0.035$.

\begin{table}[htb]
\begin{center}
\begin{tabular}{ccc}
Parameter & ~ & Value \\
\hline
$M_1$ & & 102.0 $e^{i \phi_1}$ \\
$M_2$ & & 192.0 \\
$\mu$ & & 377.5 $e^{i \phi_\mu}$ \\
$\tan \beta$ & & 10 \\
$m_{\tilde e_R},m_{\tilde \mu_R}$ & & 224.0 \\
$m_{\tilde e_L},m_{\tilde \mu_L}$ & & 264.5 \\
$m_{\tilde \nu_e}$ & & 252.4
\end{tabular}
\caption{Low-energy parameters (at the scale $M_Z$) for the SUSY scenario used.
The dimensionful parameters are in GeV.
\label{tab:1}}
\end{center}
\end{table}

CP violation in the neutralino mixing matrix arises when the phases of $M_1$
and/or $\mu$ are different from $0,\pi$. These phases generally lead to
large supersymmetrc contributions to EDMs. If the squark spectrum (which does
not play
any role in our analysis) is heavy enough, experimental limits on the neutron
and Mercury EDMs are satisfied. On the other
hand, for the values for selectron masses under consideration, the experimental
bound on the electron EDM $d_e$ imposes a severe constraint on the allowed
region in the $(\phi_1,\phi_\mu)$ plane. Using
the expressions for $d_e$ in Ref.~\cite{arnowitt}, we
find that for each $\phi_1$ between 0 and $2 \pi$ there exist two narrow
intervals for $\phi_\mu$ (one with values $\phi_\mu \sim 0$ and the other with
values $\phi_\mu \sim \pi$) in which
the neutralino and chargino contributions to $d_e$ cancel, resulting in a value
compatible with the experimental limit $d_e^\mathrm{\,exp} = (0.079 \pm 0.074)
\times 10^{-26} ~ e$ cm \cite{pdb}. (For instance, for $\phi_1 = \pi/2$ we find
$\phi_\mu \simeq -0.12$ or $\phi_\mu \simeq 3.21$.) 
Therefore, in principle it is possible to have any phase $\phi_1$,
though with a strong correlation with $\phi_\mu$.
Of course, if $\phi_1$ and $\phi_\mu$ are experimentally found to be
non-vanishing, a satisfactory explanation will be necessary for this
correlation, which apparently would be a
``fine tuning'' of their values \cite{pr3}.

\begin{figure}[htb]
\begin{center}
\epsfig{file=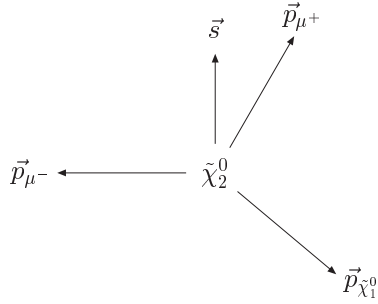,width=5cm,clip=}
\caption{Schematic picture of the $\tilde \chi_2^0$ decay in its rest frame.}
\label{fig:X2dist}
\end{center}
\end{figure}

Let us discuss which asymmetries may be defined in the $\tilde \chi_2^0$
decay. In the $\tilde \chi_2^0$ rest frame, the decay looks as depicted in
Fig.~\ref{fig:X2dist}, with $\vec s$ the $\tilde \chi_2^0$ spin and the 
3-momenta in obvious notation. Under CP, the spin and momenta transform as:
\begin{equation}
\vec s \to \vec s \,, \quad
\vec p_{\mu^+} \to -\vec p_{\mu^-} \;,\quad
\vec p_{\mu^-} \to -\vec p_{\mu^+} \;,\quad
\vec p_{\tilde \chi_1^0} \to -\vec p_{\tilde \chi_1^0} \;.
\end{equation}

Using these vectors, we can build the CP-odd quantities
\begin{eqnarray}
Q_1 & = & \vec s \cdot \left( \vec p_{\mu^-} \times \vec p_{\mu^+}
\right) \,, \nonumber \\
Q_2 & = & \vec s \cdot \left( \vec p_{\mu^-} + \vec p_{\mu^+}
\right) \,.
\end{eqnarray}
Several other CP-odd quantites can also be constructed involving higher powers
of the momenta or using the energies, e.g.\ we can define
$Q_3 = \left( E_{\mu^-} - E_{\mu^+}\right)$, which is also odd under CP. For
$Q_{1-3}$ we construct the asymmetries
\begin{equation}
A_i \equiv \frac{N(Q_i > 0) - N(Q_i < 0)}{N(Q_i > 0) + N(Q_i < 0)} \,,
\end{equation}
where $N$ denotes the number of events. Since $\tilde \chi_2^0$ and
$\tilde \chi_1^0$ are Majorana particles, these asymmetries must vanish if
CP is conserved. Hence, they are genuine signals of CP violation.
We note that $Q_{2,3}$ are even under ``naive'' time reversal T. This implies
that $A_{2,3}$ need the presence of absorptive CP-conserving phases in the
amplitude in order to be nonvanishing \cite{Teven}.
CP-conserving phases appear in loop diagrams with on-shell propagators,
through final state interactions or from the widths of intermediate unstable
particles (the phases originated by particle widths are usually tiny).
In $\tilde \chi_2^0$ decay,
T-even asymmetries like $A_{2,3}$ can result from the interference of a
dominant tree-level and a subleading loop diagram mediating the decay and are
expected to be small. On the other
hand $Q_1$ is T-odd, thus a relatively large asymmetry $A_1$ can be generated
at the tree level, without the need of an absorptive phase.
Our analysis is centred in the asymmetry $A_1$.

\section{Generation of the signals}
\label{sec:3}

We calculate the matrix element for the resonant processes in
Eqs.~(\ref{ec:proc})
using {\tt HELAS} \cite{helas}, so as to include all spin correlations and
finite width effects. The relevant terms of the Lagrangian and conventions used
can be found in Refs.~\cite{npb,next}. We assume a CM energy of 800 GeV, with 
$e^-$ polarisation $P_{e^-} = -0.8$ and $e^+$ polarisation 
$P_{e^+} = 0.6$, and an integrated luminosity of 534 fb$^{-1}$ per year
\cite{lum}. Beam polarisation does not have any effect on the asymmetries, which
are defined for $\tilde e_L$ decays independently of the production mechanism,
but increases the total signal cross sections. In our calculation we take into
account the
effects of initial state radiation (ISR) \cite{isr} and beamstrahlung
\cite{peskin,BS2}. For the latter we use the
design parameters $\Upsilon = 0.09$, $N = 1.51$ \cite{lum}.\footnote{The actual
expressions for ISR and beamstrahlung  used in our calculation can be found in
Ref.~\cite{npb}.}
We also include a beam energy spread of 1\%.

In order to simulate the calorimeter and tracking resolution, we perform a
Gaussian smearing
of the energies of electrons ($e$) and muons ($\mu$) using the 
specifications in the TESLA Technical Design Report \cite{tesla2}
\begin{equation}
\frac{\Delta E^e}{E^e} = \frac{10\%}{\sqrt{E^e}} \oplus 1 \% \;, \quad
\frac{\Delta E^\mu}{E^\mu} = 0.02 \% \, E^\mu \;,
\end{equation}
where the energies are in GeV and the two terms are added in quadrature. We
apply ``detector'' cuts on transverse momenta, $p_T \geq 10$ GeV, and
pseudorapidities $|\eta| \leq 2.5$, the latter corresponding to polar angles
$10^\circ \leq \theta \leq 170^\circ$. We also reject events in which the
leptons are not isolated, requiring a ``lego-plot'' separation
$\Delta R = \sqrt{\Delta \eta^2+\Delta \phi^2} \geq 0.4$.
We do not require specific trigger conditions, and we assume that the
presence of charged leptons with high transverse momentum will suffice.
For the Monte Carlo integration in 8-body phase space we use
{\tt RAMBO} \cite{rambo}.

The reconstruction of the final state momenta is done with the
procedure described in detail in Ref.~\cite{npb}, with some
modifications. In general, it is necessary to have as
many kinematical relations as unknown variables in order to determine the
momenta of the undetected particles. In our case, there are 8 unknowns (the 4
components of the two $\tilde \chi_1^0$ momenta) and 8 constraints. These are
derived from energy and momentum conservation (4 constraints), from the fact
that the two $\tilde \chi_1^0$ are on shell (two constraints), from the
kinematics of the decay of the $\tilde \chi_2^0$ (one constraint). The last
constraint comes
from the hypothesis that in $e^+ e^-$ collisions two particles are
produced, either with the same mass or having a squared mass difference equal
to $m_{\tilde e_L}^2 - m_{\tilde e_R}^2$.
The reconstruction is attempted for both cases, considering different effective
CM energies $E_\mathrm{eff} < E_\mathrm{CM}$ (to partially take into
account ISR and beamstrahlung effects), selecting the one which gives
reconstructed selectron masses closest to their actual values which
can be measured in other processes \cite{tdr,martyn,feng,blochinger}.
If the event does not reasonably fit into any of the two possibilities, it is
discarded.
The identification of  $\tilde e_L \tilde e_L$ versus $\tilde e_R \tilde e_L$
production is successful in most cases, with a good ``tagging'' 80\% of the time
for $\tilde e_L \tilde e_L$ and 70\% for $\tilde e_R \tilde e_L$. We note that
these processes are topologically very similar, being the only difference the
selectron energies in CM frame (and thus the magnitude of their 3-momentum).
In the former process, both selectron energies are 400 GeV, whereas in the
latter they are 412 GeV for $\tilde e_L$, 388 GeV for $\tilde e_R$.

The reconstruction procedure determines the momenta of the two
unobserved $\tilde \chi_1^0$,  identifying for $\tilde e_L^+ \tilde e_L^-$
whether the selectron pair
has decayed in the channel $\tilde e_L^+ \to e^+ \tilde \chi_1^0$,
$\tilde e_L^- \to e^- \tilde \chi_2^0 \to e^- \tilde \chi_1^0 \mu^+ \mu^-$
or, on the contrary, in the channel
$\tilde e_L^+ \to e^+ \tilde \chi_2^0 \to e^+ \tilde \chi_1^0 \mu^+ \mu^-$,
$\tilde e_L^- \to e^- \tilde \chi_1^0$. The knowledge of all
final state momenta, as well as the identification of the particles
resulting from each decay, allows us to construct various mass, angular and
energy distributions \cite{note}, and in particular the determination of
the selectron and $\tilde \chi_2^0$ rest frames.
In each event, the $\tilde \chi_2^0$ spin direction $\vec s$ can be found
as follows: If the $\tilde \chi_2^0$ results from the decay of a $\tilde e_L^-$,
it has negative helicity, so its spin direction is $\vec s=-\vec p$, with $\vec
p$ the $\tilde \chi_2^0$ momentum in the $\tilde e_L^-$ rest frame.
If the $\tilde \chi_2^0$ results from the decay of a $\tilde e_L^+$, it has
positive
helicity and its spin direction is $\vec s=\vec p$, with $\vec
p$ the $\tilde \chi_2^0$ momentum in the $\tilde e_L^+$ rest frame.

Finally, it is necessary to discuss the possible SM and SUSY backgrounds to our
signal. The relevant SUSY backgrounds are sneutrino and $\tilde \chi_2^0$
pair production, in the decay channels
\begin{align}
e^+ e^- & \to \tilde \nu_e^* \tilde \nu_e \to e^+ \tilde \chi_1^- \, e^- \tilde
\chi_1^+ \to e^+ \tilde \chi_1^0 \mu^- \bar \nu_\mu  \, e^- \tilde \chi_1^0
\mu^+ \nu_\mu  \,, \nonumber \\
e^+ e^- & \to \tilde \nu_e^* \tilde \nu_e \to e^+ \tilde \chi_1^- \, \nu_e \,
\tilde \chi_2^0 \to e^+ \tilde \chi_1^0 e^- \bar \nu_e \, \nu_e \,
\tilde \chi_1^0 \mu^+ \mu^- \,, \nonumber \\
e^+ e^- & \to \tilde \chi_2^0 \tilde \chi_2^0 \to \tilde \chi_1^0 e^+ e^-
\, \tilde \chi_1^0 \mu^+ \mu^- \,.
\label{ec:EEb}
\end{align}
In sneutrino pair production the $\tilde \chi_1^-  \tilde \chi_1^+$ decay
channel has a larger branching ratio, with
$\mathrm{Br}(\tilde \nu \to e^- \tilde \chi_1^+) = 0.52$,
$\mathrm{Br}(\tilde \chi_1^+ \to \tilde \chi_1^0 e^+ \nu ) = 0.1$,
while for the the $\tilde \chi_1^-  \tilde \chi_2^0$ mode
$\mathrm{Br}(\tilde \nu \to \nu \tilde \chi_2^0) = 0.22$,
$\mathrm{Br}(\tilde \chi_2^0 \to \tilde \chi_1^0 \mu^+ \mu^-) = 0.035$.
We have calculated the three processes in the same way described for selectron
pair production, and include them in our results. The
cross sections of these backgrounds turn out to be 4 times larger than the
signals, but can be reduced with the reconstruction of the final state
momenta.
The reconstruction method applied for the signal partially eliminates the three
backgrounds. The first and third ones are further reduced applying a
reconstruction
procedure specific for each case (not discussed here for brevity) and requiring
that signal and background events do
not have a kinematics similar to sneutrino or $\tilde \chi_2^0$ pair
production.\footnote{In the
$\tilde \chi_1^-  \tilde \chi_1^+$ channel only the momenta of each $\nu
\tilde \chi_1^0$ pair can be determined, but that is enough to obtain the
chargino and sneutrino momenta.}
This is not possible for the second background in Eq.~(\ref{ec:EEb}) due to the
different kinematics of the process.
A cut requiring transverse energy $H_T > 200$ GeV is applied as
well. Other SUSY backgrounds like $e^+ e^- \to \tilde \chi_1^\pm \chi_2^\mp
\to \chi_1^+ \chi_1^- Z \to e^+ e^- \mu^+ \mu^- \nu \bar \nu \tilde \chi_1^0
\tilde \chi_1^0$ (involving several decay channels which lead to the same
final state of $e^+ e^- \mu^+ \mu^-$ plus missing energy and momentum)
are much smaller, with cross sections smaller than 0.01 fb.
The SM background is given by six fermion production $e^+ e^- \to
e^+ e^- \mu^+ \mu^- \nu \bar \nu$. Its cross section is below 0.3 fb
\cite{lusifer} and with kinematical cuts can be eliminated more easily than SUSY
backgrounds.

\section{Numerical results}
\label{sec:4}

\begin{figure}[htb]
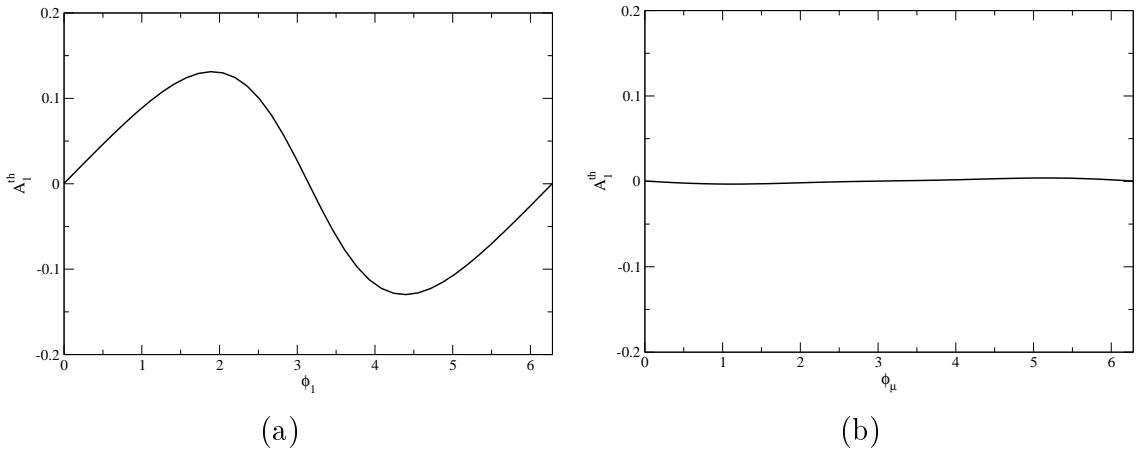

\begin{center}
\begin{tabular}{cc}
\epsfig{file=Figs/asim-th.eps,width=7.3cm,clip=} &
\epsfig{file=Figs/asim-mu-th.eps,width=7.3cm,clip=} \\
(a) & (b)
\end{tabular}
\caption{Theoretical value of $A_1$ as a function of $\phi_1$, for $\phi_\mu =0$
(a) and as a function of $\phi_\mu$, for $\phi_1=0$ (b).}
\label{fig:A1th}
\end{center}
\end{figure}

\begin{figure}[htb]
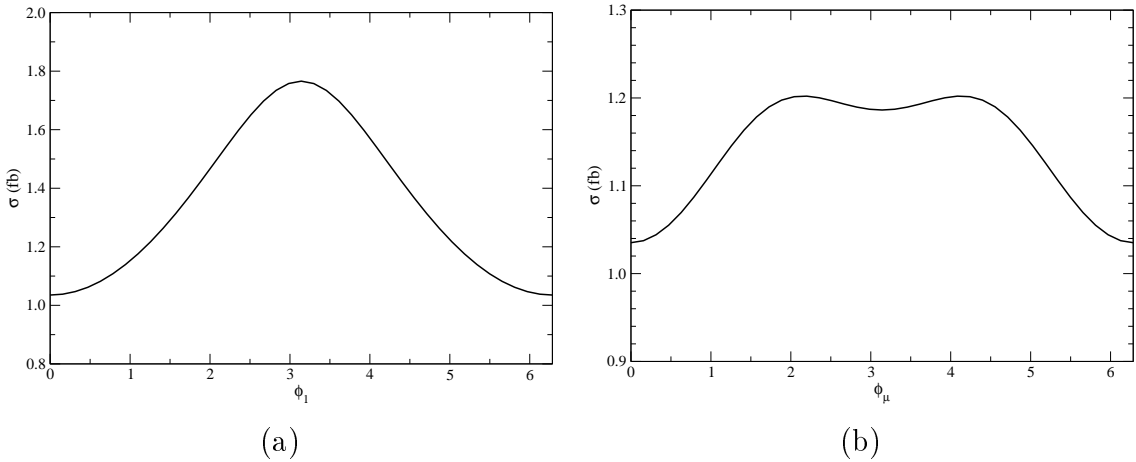

\begin{center}
\begin{tabular}{cc}
\epsfig{file=Figs/cross-th.eps,width=7.3cm,clip=} &
\epsfig{file=Figs/cross-mu-th.eps,width=7.3cm,clip=} \\
(a) & (b)
\end{tabular}
\caption{Total cross section (without ISR, beamstrahlung and beam spread
effects, and without any kinematical cuts) as a function of $\phi_1$, for
$\phi_\mu =0$ (a) and as a function of $\phi_\mu$, for $\phi_1=0$ (b).}
\label{fig:sigth}
\end{center}
\end{figure}

We first examine the possible values that this asymmetry may take in
connection with
EDM constraints. In Fig.~\ref{fig:A1th} we show its dependence on the
two phases $\phi_1$, $\phi_\mu$ (these plots are calculated with the Monte Carlo
described in last section, but assuming perfect momenta reconstruction and
particle identification, and without any kinematical nor detector cuts).
For $\phi_1$ sufficiently large $A_1$ reaches values of $\pm 0.13$,
while for $\phi_1=0$ the asymmetry is negligible independently of
$\phi_\mu$.\footnote{The main contribution to the asymmetry comes from the
interference between the $Z$ diagram in Fig.~\ref{fig:X2decay}a and the
diagrams with $\mu_R$ exchange in Fig.~\ref{fig:X2decay}c and
\ref{fig:X2decay}d. 
We note that the $Z$ contribution alone produces an
asymmetry $A_1 = 0.021$ without the need of interference, but the asymmetry is
much larger when the rest of diagrams are included.}
The dependence of the cross section on these two phases is plotted
in Fig.~\ref{fig:sigth}.
Our approach is then as follows: for each value of $\phi_1$, we know from
the discussion in Section~\ref{sec:2} that there is an allowed
interval of $\phi_\mu$ (which we may take with $|\phi_\mu| \lesssim 0.12$)
in which the electron EDM does not exceed the experimental bound.
We then calculate $A_1$ for this $\phi_1$, but taking $\phi_\mu = 0$, since the
asymmetry is almost independent of the latter phase and the effect on the cross
section is also rather small for $|\phi_\mu| \lesssim 0.12$.

\begin{figure}[htb]
\begin{center}
\epsfig{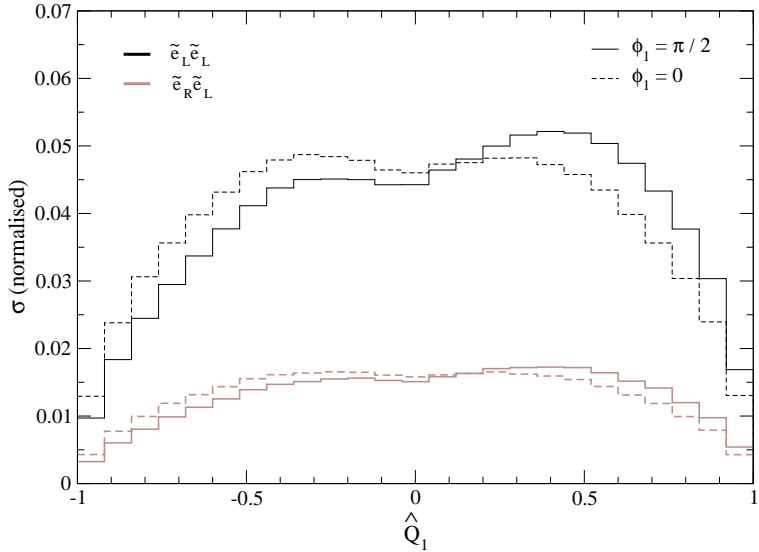}
\caption{Kinematical distribution of the normalised triple product $\hat Q_1$.}
\label{fig:Q1}
\end{center}
\end{figure}

The cross sections for $\phi_1 = 0$ of the different processes in
Eqs.~(\ref{ec:proc},\ref{ec:EEb}) are collected in
Table~\ref{tab:cross}, before and after final state momenta reconstruction and
kinematical cuts. All figures include
the various corrections discussed in the previous section. The
reconstruction of the signal and backgrounds allows us to reduce the latter by a
factor of 20, while keeping 80\% of the signal.
We define the normalised quantity
$\hat Q_1 = \vec s \cdot \left( \hat p_{\mu^-} \times \hat p_{\mu^+} \right)$,
using unit vectors in the direction of the muon momenta, so that $-1 \leq \hat
Q_1 \leq 1$. The kinematical distribution of $\hat Q_1$ for the signals
is shown in Fig.~\ref{fig:Q1}, taking $\phi_1 = 0$ and $\phi_1 = \pi/2$. We have
normalised the $\tilde e_L^+ \tilde e_L^-$ cross sections to unity, and the
$\tilde e_R^\pm \tilde e_L^\mp$ ones to $1/3$. In this plot we take into account
ISR, beamstrahlung, beam spread and detector effects.
The $\hat Q_1$ distribution already shows
that the asymmetry is a real dynamical effect and is not
a fake asymmetry caused by ``detector'' cuts applied in phase space. In our
calculation we find that for $\phi_1 = \pi/2$ the relative difference between
the phase space volumes of the two hemispheres
(with $\hat Q_1 < 0$ and $\hat Q_1 > 0$) is negligible, of $3.1 \times 10^{-4}$
for $\tilde e_L^+ \tilde e_L^-$ and $6.9 \times 10^{-4}$ for $\tilde e_R^\pm
\tilde e_L^\mp$. We note that this is not the case for the
asymmetry $A_2$. Even in the CP-conserving case with $\phi_1=0$ a
fake asymmetry $A_2 = -0.055$ is generated just by phase space cuts, and might
obscure the observation of a real CP asymmetry, which is expected to be very
small in this case.

\begin{table}[htb]
\begin{center}
\begin{tabular}{ccc}
 & Before & After \\
\hline
$\tilde e_L^+ \tilde e_L^-$ & 0.21 & 0.18 \\
$\tilde e_R^+ \tilde e_L^-$ & 0.061 & 0.046 \\
$\tilde e_L^+ \tilde e_R^-$ & 0.027 & 0.021 \\
$\tilde \nu^* \tilde \nu$ ($\tilde \chi_1^-  \tilde \chi_1^+$) & 0.99 & 0.039 \\
$\tilde \nu^* \tilde \nu$ ($\tilde \chi_1^-  \tilde \chi_2^0$) & 0.24 & 0.011 \\
$\tilde \chi_2^0 \tilde \chi_2^0$ & 0.11 & 0.0081 \\
\end{tabular}
\end{center}
\label{tab:cross}
\caption{Cross sections (in fb) before and after signal reconstruction and
kinematical cuts of the processes in Eqs.~(\ref{ec:proc},\ref{ec:EEb}).}
\end{table}

\begin{figure}[ht]
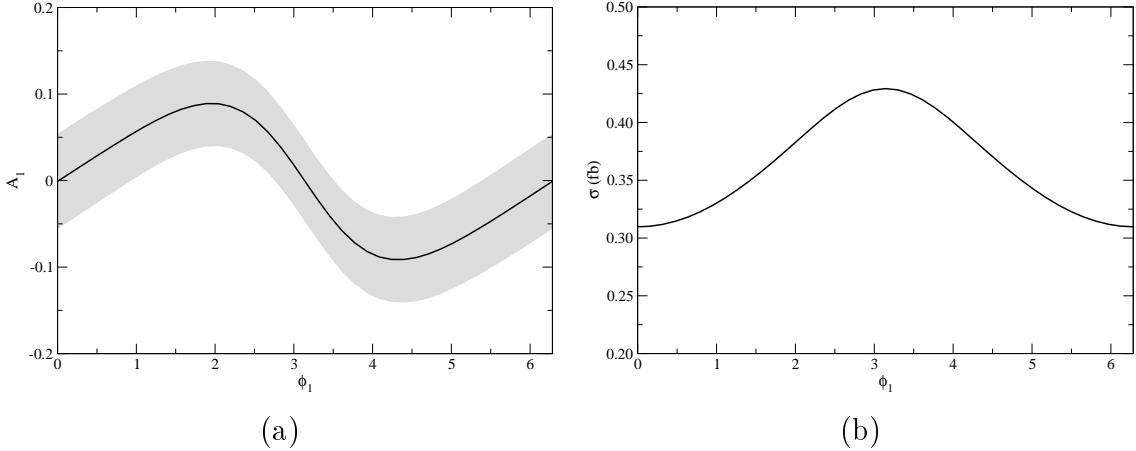

\begin{center}
\begin{tabular}{cc}
\epsfig{file=Figs/asim.eps,width=7.3cm,clip=} &
\epsfig{file=Figs/cross.eps,width=7.3cm,clip=} \\
(a) & (b)
\end{tabular}
\caption{(a) Asymmetry $A_1$ as a function of the phase $\phi_1$. The shaded
region
represents the statistical error for two years of running. (b) Total cross
section for $e^+ e^- \to \tilde e_{L,R}^\pm \tilde e_L^\mp
\to e^+ e^- \mu^+ \mu^- \tilde \chi_1^0 \tilde \chi_1^0$ as a function of
$\phi_1$. In both cases the backgrounds from $\tilde \nu^* \tilde \nu$ and
$\tilde \chi_2^0 \tilde \chi_2^0$ production are included.}
\label{fig:total}
\end{center}
\end{figure}

The difference between the distributions for the two values of the phase
is smaller for the $\tilde e_R \tilde e_L$ signals. This is because the
reconstruction procedure has a smaller efficiency for the correct identification
of $\tilde e_R^\pm \tilde e_L^\mp$ events,
leading to a small washing-out of the asymmetry. For $\phi_1 = \pi/2$, we have
for $\tilde e_L^+ \tilde e_L^-$ the value $A_1 = 0.110$, while for
$\tilde e_R^+ \tilde e_L^-$ and $\tilde e_R^- \tilde e_L^+$
we find $A_1 = 0.098$, $A_1 = 0.099$, respectively. 
The total asymmetry $A_1$ as a function of $\phi_1$ is plotted
in Fig.~\ref{fig:total}a, including the backgrounds in Eqs.~(\ref{ec:EEb}),
which have vanishing CP asymmetry, as well as ISR, beamstrahlung, beam spread
and detector effects.
The shaded region represents the statistical error
for two years of running, with a luminosity of 534 fb$^{-1}$ per year.
The cross section also exhibits a strong dependence on $\phi_1$, as can be
observed in Fig.~\ref{fig:total}b. Nevertheless, the measurement of the cross
section is not likely to provide any information on the phase $\phi_1$, due to
the multiple theoretical uncertainties present regarding the neutralino
mixing matrix, sparticle masses, scale dependence, etc.

\section{Summary}
\label{sec:6}

In this work we have shown that a CP asymmetry in selectron cascade decays
$\tilde e_L \to e \tilde \chi_2^0 \to e \tilde \chi_1^0 \mu^+ \mu^-$ can be
observable in SUSY scenarios where $\tilde \chi_2^0$ has three-body decays,
if there is a large nonzero phase $\phi_1$ in the gaugino mass $M_1$. The key
features for the observation of the CP asymmetry are: ({\em i\/}) the neutralinos
produced from selectron decays are 100\% polarised; ({\em ii\/}) all final
state momenta can be accurately reconstructed in the processes under
consideration; ({\em iii\/}) this reconstruction allows us to eliminate
potentially dangerous backgrounds.

In SUSY scenarios with three-body decays of $\tilde \chi_2^0$, a
sizeable $\tilde e_L$ production
can only take place at a TESLA upgrade with a CM energy of 800 GeV.
We have selected one of such scenarios, with $\tilde \chi_1^0$ and $\tilde
\chi_2^0$ gaugino-like, and shown that the CP asymmetry in
the triple product $\vec s \cdot (\vec p_{\mu^-} \times \vec p_{\mu^+})$ could
be up to $A_1 = \pm 0.1$ for large phases $\phi_1 \simeq \pm 2$. These
asymmetries could be observable with 1.8 standard deviations after two years of
running with the proposed luminosity. It can also be seen that
in similar scenarios where $\tilde \chi_1^0$ and $\tilde \chi_2^0$ have larger
Higgsino components the asymmetries could be even larger, and observable as
well. The results here can be compared with the CP asymmetry in
$\tilde \chi_1^0 \tilde \chi_2^0$ production \cite{bartl2} in the same SUSY
scenario. This analysis will be presented elsewhere \cite{next}. In $\tilde
\chi_1^0 \tilde \chi_2^0$ production, although the cross section is much higher,
the asymmetry
is smaller and there are large backgrounds that further reduce it. The maximum
statistical significance obtained for the asymmetry is $1.5 \, \sigma$
when $\phi_1 = 3 \pi/4$.
At any rate these two processes are complementary, because the
asymmetries have a different dependence on $\phi_1$ in each case.
For instance, the CP asymmetry in $\tilde \chi_1^0 \tilde \chi_2^0$ production
almost vanishes for $\phi_1 = \pi/2$ and $\phi_1 = 3 \pi/2$ \cite{next},
while in selectron cascade decays it is nearly maximal.

To conclude, we stress that the existence of CP-violating phases in
SUSY-breaking terms of the Lagrangian is still an open question. In this paper
we have focused on the phases $\phi_1$ and $\phi_\mu$ of the parameters
$M_1$ and $\mu$, respectively. Assuming a relatively light SUSY spectrum there
are two alternative possibilities to explain the unobserved EDM of the
electron:
either $M_1$ and $\mu$ are approximately real or on the contrary they have
large phases and there exist
cancellations between neutralino and chargino contributions to the electron
EDM. In this situation, the study of observables with a different functional
dependence on  $\phi_1$ and $\phi_\mu$, as the CP asymmetry investigated here,
is of great help in order to confirm one of the two hypotheses.

\vspace{1cm}
\noindent
{\Large \bf Acknowledgements}

\vspace{0.4cm} \noindent
I thank A. Bartl, S. Hesselbach and A. M. Teixeira for collaboration at the
initial stage of this work. This work has been supported
by the European Community's Human Potential Programme under contract
HTRN--CT--2000--00149 Physics at Colliders and by FCT
through projects POCTI/FNU/43793/2002, CFIF--Plurianual (2/91) and
grant SFRH/BPD/12603/2003.


\end{document}